\documentclass[11pt,a4paper]{article}
\usepackage{jcappub}

\def\be{\begin{equation}}
\def\ee{\end{equation}}
\def\beq{\begin{equation}}
\def\eeq{\end{equation}}
\def\bea{\begin{eqnarray}}
\def\eea{\end{eqnarray}}
\def\bml{\begin{subequations}}
\def\blea{\bml\begin{eqnarray}}
\def\elea{\end{eqnarray}\end{subequations}}

\begin{document}

\title{Dynamical systems of eternal inflation:
\newline
A possible solution to the problems of entropy, measure, observables and initial conditions}

\author{Vitaly Vanchurin}

\emailAdd{vanchurin@stanford.edu}

\date{\today}

\affiliation{Stanford Institute of Theoretical Physics and Department of Physics, Stanford University, Stanford, CA 94305, USA}

\abstract{
There are two main approaches to non-equlibrium statistical mechanics: one using stochastic processes and the other using dynamical systems. To model the dynamics during inflation one usually adopts a stochastic description, which is known to suffer from serious conceptual problems. To overcome the problems and/or to gain more insight, we develop a dynamical systems approach. A key assumption that goes into analysis is the chaotic hypothesis, which is a natural generalization of the ergodic hypothesis to non-Hamiltonian systems. The unfamiliar feature for gravitational systems is that the local phase space trajectories can either reproduce or escape due to the presence of cosmological and black hole horizons.  We argue that the effect of horizons can be studied using dynamical systems  and  apply the so-called thermodynamic formalism to derive the equilibrium (or Sinai-Ruelle-Bowen) measure given by a variational principle. We show that the only physical measure is not the Liouville measure (i.e. no entropy problem), but the equilibrium measure (i.e. no measure problem) defined over local trajectories (i.e. no problem of observables) and supported on only infinite trajectories (i.e. no problem of initial conditions). Phenomenological aspects of the fluctuation theorem are discussed. } 

\maketitle

\section{Introduction}

How to make testable and sensible prediction is one of the most important unresolved problems in contemporary cosmology. A number of interesting, but controversial, ideas had been put forward (e.g. quantum cosmology \cite{QuantumCosmology, HartleQuantum}, holographic cosmology \cite{DSCFT, FRWCFT, DSDS, HST, HVG}), but by far the most popular approach is realized in the context of eternal inflation \cite{EternalSteinhardt, EternalVilenkin, EternalLinde}, where the problem of making predictions is known as the measure problem \cite{LindeStochastic,VilenkinStochastic, StationaryStochastic}. In recent years the idea of eternal inflation has gained a renewed interest due to a possible unification of inflationary cosmology and string theory in the context of a huge landscape of vacua \cite{Landscape}. It is  also argued that the unified framework may simultaneously help us to solve the cosmological constant problem using either a non-anthropic solution \cite{NonAnthropics} or an anthropic solution \cite{Anthropics} with very mild assumptions about an underlying probability measure. However, to declare a victory one has to derive the measure from first principles which has proven to be a very difficult task. So, the main question is:  can the measure problem in eternal inflation be really solved? 

The answer, perhaps, depends crucially on how one defines inflation. So far, most of the attempts to tackle the problem were using stochastic description which can be modeled, for example, by diffusion in a configuration space. Given the stochastic model one can start asking probabilistic questions, but, as it turned out, the answer always depends on either initial conditions \cite{StarobinskyStochastic, LocalStochastic, BoussoStochastic} (i.e. {\it problem of initial conditions}) or on a cut-off procedure \cite{LindeStochastic, VilenkinStochastic, StationaryStochastic} (i.e. {\it measure problem}). This would be a real pity if one had to postulate an additional rule such as initial conditions or a probability measure to determine observables in a system which seems to have an attractor dynamics (e.g. cosmic inflation). Moreover, many otherwise phenomenologically acceptable stochastic measures (e.g. causal patch measure \cite{BoussoStochastic} or scale factor measure \cite{ScaleFactor}) give rise to very counterintuitive and somewhat paradoxical predictions \cite{GeocentricCosmology,GuthVanchurin,  Paradoxes}. In other words, it is not always clear how to define a probability space of observables without violating the basic principles of the probability theory. We will refer to it as the problem of defining cosmological observables or simply the {\it problem of observables}.  At this point one might start worrying whether the stochastic description, which is at most an approximation to the underlying microscopic dynamics, is a good mathematical model of eternal inflation. The objective of this paper is to construct an alternative mathematical model of inflation using dynamical systems, but before we proceed, let us briefly review another related problem - the {\it entropy problem} \cite{EntropyProblem}.

Consider a finite Hamiltonian system. For such systems the most physical time-invariant measure is given by the Liouville measure, according to which a typical observer should find himself in a highly entropic state. Cosmology for such observers (often called Boltzmann observers \cite{BoltzmannBrains}) would be very boring in a sharp contrast to what we actually observe. This is the so-called entropy problem. On the other hand, one can certainly define other non-invariant measures on the surface of initial conditions of a given observer (e.g. geocentric measures \cite{GeocentricCosmology}), but it might be more desirable to have a dynamical mechanism to explain cosmological observations.\footnote{Indeed, in its origin the geocentric approach adopts a Bayesian (or subjective) interpretation of probabilities when a frequentists (or objective) interpretation is often easier digested by physicists.} So, another relevant question is: can the entropy problem in cosmology be really solved? 

There are at least two approaches that one might take. Roughly speaking, we need to violate the Liouville theorem by providing a mechanism to either add or remove phase space trajectories. Clearly, a global description of gravitational systems provides a natural mechanism to accommodate both phenomena. For example, the eternally inflating space-time constantly ``adds'' new local trajectories (i.e. more and more local observers fall out of causal contact with each other) and the constantly forming black holes ``remove'' old local trajectories when the observers hit singularities. Since the phase space volume for such observers is no longer conserved, the Liouville measure is not very useful, but one might still wonder whether there are any good time-invariant measures. As we will argue below, the space of time-invariant measures for a generic dynamical system is very large, but the so-called equilibrium measure is often a unique measure given by a variational principle. So, it appears that the entropy and measure problems can be simultaneously avoided, if not solved, in the context of more general dynamical systems. In addition, the equilibrium measure is defined on a space of local trajectories with support on only infinite trajectories.\footnote{Although the precise mathematical definition of equilibrium measures involves an infinite time limit, for all practical purposes it is sufficient to follow the system for a very long but finite time. Moreover, for the systems which only allow finite trajectories the phrase ``infinite trajectories'' should be read throughout the paper as ``very long trajectories''.} This provides a possible resolution to the problem of defining relevant cosmological observables (i.e. observables are the trajectories) as well as to the problem of initial conditions (i.e. for infinite trajectories the problem is irrelevant). 

The paper is organized as follows. In Section \ref{SecStochastic}  we review the stochastic approach and problems associated with it. In Section \ref{SecDynamical} we introduce the dynamical systems approach with an emphasis to a variational principle and a fluctuation theorem. In Section \ref{SecEternal} we construct a dynamical system of eternal inflation and derive its equilibrium measures. In Section \ref{SecResults} we summaries the main results.

\section{Stochastic approach}\label{SecStochastic}

Consider a deterministic dynamical system whose evolution is defined by a velocity flow ${\bf v}({\bf x}) \equiv  \frac{d \bf x}{d t}$ and the system at time $t$ is described by a state vector ${\bf x}(t) \in X$. If the system does not have any absorbing states, then the evolution of an arbitrary distribution function $\mu({\bf x},t)$ can be followed in time using the continuity equation:
\be
\frac{\partial \mu({\bf x},t)}{\partial t} = - \frac{\partial}{\partial {\bf x}} \cdot  {\bf v}({\bf x}) \mu({\bf x},t).
\label{Eq:Continuity}
\ee
The main challenge in the stochastic approach is to solve the continuity equation for a given model of the velocity flow ${\bf v}({\bf x})$. In what follows we will consider three models of the flow all of which lead to yet unresolved cosmological problems. 

\subsection{Entropy problem}\label{PENT}

Perhaps the most studied dynamical systems are Hamiltonian systems. For such systems the components of a state vector ${\bf x}$ come in conjugate pairs ${\bf x} =\{{\bf p, q}\}$ corresponding to momentum ${\bf p}$  and position ${\bf q}$ coordinates and the Hamiltonian equations of motion imply that the velocity flow is not compressible $ \frac{\partial}{\partial {\bf x}} \cdot  {\bf v} = 0$. Under the incompressibility assumption Eq. (\ref{Eq:Continuity}) becomes the classical Liouville equation:
\be
\frac{\partial \mu({\bf x},t)}{\partial t} = -  {\bf v}({\bf x})   \cdot   \frac{\partial}{\partial {\bf x}} \mu({\bf x},t),
\ee 
which has a trivial time-independent solution $\mu_L = \Gamma^{-1}$ known as the Liouville measure, where $\Gamma$  is the volume of phase space $X$. This measure is known to  be very useful for describing large thermodynamic system,  but it is not very useful for describing the universe. In particular, the Liouville measure gives rise to the entropy problem - entropy of the observable universe is much smaller than what one would naively expect \cite{EntropyProblem}. It also follows immediately that for infinite  ($\Gamma=\infty$) Hamiltonian systems the Liouville measure does not exist, which is the main source of a measure problem to be discussed below. 

But what is really the size of the phase space of eternal inflation: finite or infinite? In a flat slicing of de Sitter space,
\be 
d s^2 = dt^2 - e^{2 H t} dx^2,
\ee
the new degrees of freedom constantly come from under the Planck scale, but as the modes are eventually stretched out to super-horizon scales they can no longer be observed. This is a global picture that suggests that the total number of degrees of freedom is infinite, which is misleading if one wants to count only the states accessible to a local observer. From a local viewpoint, the evolution is most conveniently described in static coordinates,
\be 
d s^2 = (1-(r H)^2) dt^2 - \frac{d r^2}{1-(r H)^2}  - r^2 d \Omega, 
\ee
where the amount of  information accessible to a local observer is only finite (when a cut-off is imposed at the Planck scale). Note that in a quasi-de Sitter space, with possible transitions between different vacua, the size of the phase space should be set by an exponent of the entropy of a vacua with with the smallest positive energy density. One could still ran into problems with Minkowski vacua, but this will turn out not to be the case for the time-invariant measures discussed in Section \ref{SecEternal}. The phase space might still be huge but, what is more important, it is finite.

\subsection{Measure problem}\label{PMSR}

If we try to model a finite dynamical system, consisting of only degrees of freedom accessible to a local observer, then to solve the entropy problem one should abandon the idea of a  Hamiltonian description whose predictions are in conflict with observations. The appearance of a non-Hamiltonian dynamics is not entirely new and happens all the time whenever some of the degrees of freedom of a larger Hamiltonian system are ignored. This is exactly the situation in a local description of gravitational systems with horizons. Once we integrate over the degrees of freedom unaccessible to a local observer, the local dynamical system should start to behave as a non-Hamiltonian system. 

In a global stochastic description of eternal inflation one usually models the dynamics with three non-Hamiltonian ingredients \cite{LindeStochastic}. First of all, the quantum (or thermal) effects are modeled by a compressible flow in configuration space (e.g. $\frac{1}{8 \pi^2} \partial_{\bf q}  H({\bf q})^{\gamma} \partial_{\bf q}  H({\bf q})^{3-\gamma} \mu({\bf q},t)$ during slow-roll inflation). Moreover, the constant addition and removal of local trajectories are modeled with reproducing (e.g. $3 H({\bf q})^\alpha \mu({\bf q},t)$) and absorbing states (e.g. $\mu({\bf q}) = 0$ for ${\bf q} \in \partial X$). Then, Eq. (\ref{Eq:Continuity}) becomes a branching-diffusion equation with escape:
\be
\frac{\partial \mu({\bf q},t)}{\partial t} = \frac{1}{8 \pi^2} \frac{\partial}{\partial {\bf q}} H({\bf q})^{\gamma} \frac{\partial}{\partial {\bf q}}  {H({\bf q})^{3-\gamma}} \mu({\bf q},t) - {\bf v}({\bf q})  \cdot  \frac{\partial}{\partial {\bf q}}  \mu({\bf q},t) + 3 H({\bf q})^\alpha \mu({\bf q},t),
\label{Eq:GlobalStochastic}
\ee 
where $H({\bf q})$ is the Hubble scale. 

Although the stochastic eternal inflation avoids the entropy problem, it immediately introduces a well known measure problem, which one can think of as a counterpart of the entropy problem for more general dynamical systems. In other words, what time coordinate (or $\alpha$) should we use for calculating probabilities? It is well known that different choices can lead to very different answers, and the most popular choice $\alpha=0$, corresponding to the scale factor measure, is often chosen on purely phenomenological grounds  \cite{ScaleFactor}. This might be acceptable phenomenologically, but it is not acceptable from the theoretical viewpoint where one wants to derive the measure from first principles. 

\subsection{Problem of initial conditions}\label{PICS}

Since the main source of the measure problem was due to the presence of reproducing states one might wonder what would happen if we ignore the reproduction term, $3 H({\bf x})^\alpha \mu({\bf x},t)$. In fact, this is a well known limit, corresponding to a local description of eternal inflation, where one concentrates on the evolution of comoving distributions, i.e.
\be
\frac{\partial \mu({\bf q},t)}{\partial t} = \frac{1}{8 \pi^2} \frac{\partial}{\partial {\bf q}} H({\bf q})^{\gamma}\frac{\partial}{\partial {\bf q}}  {H({\bf q})^{3-\gamma}} \mu({\bf q},t) - {\bf v}({\bf q})  \cdot  \frac{\partial}{\partial {\bf q}}  \mu({\bf q},t).
\label{Eq:LocalStochastic}
\ee 
The local stochastic approach was originally proposed to study the effects of quantum fluctuations during inflation  \cite{EternalVilenkin, StarobinskyStochastic}, but later it was adopted to study the landscape models of eternal inflation \cite{LocalStochastic}. Due to the presence of absorbing states the answer always depends on the initial conditions and even unequal weighting (e.g. entropic or anthropic) of states would not cure the problem. In other words, even if a given local measure (e.g. causal patch measure \cite{BoussoStochastic}) gives phenomenologically acceptable results for some ranges of initial conditions and some ranges of parameters of a model, it does not solve the problem of initial conditions.

One can certainly take a point of view that any physical problem must involve the knowledge of initial conditions. This was an attitude in the early days of quantum cosmology \cite{QuantumCosmology} as well as very recently in the geocentric approach \cite{GeocentricCosmology}. But then it seems unnecessary complicated to postulate a measure in addition to postulating initial conditions. If the only role of cosmology is to assign probabilities to local observations, then it is always only a problem of initial conditions and instead of fighting it we should learn how to construct a theory of initial conditions \cite{QuantumCosmology,GeocentricCosmology}. 

This would have been an acceptable ``solution'' if we did not have examples where, in the long run, the system completely forgets its initial state. For example, a large Hamiltonian system close to a thermal equilibrium is a system for which one can study its macroscopic properties without the knowledge of initial conditions. Of course, as we have argued above,  the universe is not in a thermal state, but one might still hope that a similar phenomena would occur for more general, and perhaps, non-Hamiltonian systems. In constructing a system which eventually forgets its initial state we should be careful not to introduce any other problems as it was in the case of a global description of eternal inflation discussed above.

However, in our opinion, the best possible solution to the problem of initial conditions would be if the initial conditions did not exist. In other words, if the universe would be infinite to the past (as well as to the future), then the question of initial condition would be irrelevant. For some time it was believed that eternal inflation might provide a possible framework to accomplish this task, until a no-go theorem was proved which states that the eternally inflating space-times are not past complete \cite{BGV}.  Of course, to prove any no-go theorem one makes certain assumptions which often turn out to be false and finding such ``loopholes'' is one of the biggest challenges for theoretical physics. In fact, the conclusions of Ref. \cite{BGV} do not apply to infinite trajectories, that play a central role in the dynamical systems approach developed in this paper, even when their Liouville measure is zero.

\subsection{Problem of observables}\label{POBS}

Another problem associated with stochastic descriptions of eternal inflation is related to the problem of defining relevant cosmological observables. More precisely, the problem is to define a measurable space of observables on which the cosmological measures are to be constructed. A priori, there is a lot of freedom in choosing a relevant measurable space and some popular choices include a space of local states, a space of states of local observers, or even a space of states of local brains. However, it turned out to be a very non-trivial task to define a measurable space which avoids paradoxes \cite{LindeStochastic, Youngness, GeocentricCosmology, GuthVanchurin, Paradoxes}. 

For example, if one applies the (global or local) stochastic measures to laboratory experiments, then even the most popular phenomenological choices (scale-factor measure \cite{ScaleFactor} and causal patch measure \cite{BoussoStochastic}) are not free of logical inconsistencies \cite{GeocentricCosmology, GuthVanchurin, Paradoxes}. The problems arise due to an exponential growth of the distribution $\mu({\bf x},t)$ defined on a measurable space of local states. In such exponentially growing models one can construct paradoxical situations where the probabilities of past events change with time  \cite{GuthVanchurin}. This is, perhaps, an indication that the measure $\mu({\bf x},t)$ on a space of local states might not be suitable for describing inflationary systems. 

For Hamiltonian systems the measure on states was certainly very useful for calculating macroscopic observables using microcanonical, canonical or grand canonical ensembles, but it does not have to be appropriate for more general dynamical systems. The measure $\mu({\bf x},t)$ contains only a very limited amount of information about the dynamics which was sufficient for equilibrium statistical mechanics, but might be insufficient for describing eternal inflation. For example, if the relevant distributions are to be defined on a space of trajectories then such distributions would contain much more information than any distribution on states. Evidently, one can easily calculate a measure on states from a measure on trajectories by using a sequential cutoff measure (see Ref. \cite{GuthVanchurin} for details), but not the other way around. 

\section{Dynamical systems approach}\label{SecDynamical}

A stochastic approach to statistical mechanics was originated over a century ago by Boltzmann, while a dynamical systems approach was proposed by Ruelle only forty years ago and later developed into a consistent mathematical framework \cite{Reviews}. Although most of the precise results are known only for mathematically ``simple'' systems such as Anosov (or hyperbolic) systems, the more complicated dynamical systems are usually analyzed under the so-called chaotic hypothesis. It says that for computing macroscopic observables, any chaotic dynamical system can be considered as an Anosov system. In contrast to non-chaotic (or integrable) systems, the chaotic systems allow us to define time averages independent of initial conditions which is a desired property if one wants to solve the cosmological problem of initial conditions. The chaotic hypothesis can be viewed as a generalization of the ergodic hypothesis to more general non-Hamiltonian systems. 

\subsection{Equilibrium measures}

The problems of interest in the measure theoretic discussions of non-Hamiltonian systems involve finding the most physical measure $\mu \in {\cal M}$, where ${\cal M}$ is the space of all time-invariant measures (i.e. measures which are invariant under the time evolution). Although the space is very large, there is often a unique measure $\mu_+$ defined as a late time attractor starting from an arbitrary  (continuous with respect to $\mu_L$) distribution.\footnote{For Hamiltonian systems $\mu_+ = \mu_L$, but for non-Hamiltonian systems the two measures need not be the same.} More precisely, if ${\cal O}({\bf x})$ is some observable (continuous with respect to $\mu_L$), then
\be
\int {\cal O}({\bf y})  \mu_+({\bf y}) d{\bf y} = \lim_{t\rightarrow\infty} \frac{1}{t} \int_{0}^{t} {\cal O}({\bf x}(\tau)) d\tau
\ee
should be satisfied almost surely for all but a measure zero of initial states ${\bf x}(0)$ with respect to $\mu_+$. These measures were originally proposed by Sinai \cite{Sinai}, Ruelle \cite{Ruelle} and Bowen \cite{Bowen} and go by the name of SRB measures \cite{Reviews}. In this article we will refer to them as the equilibrium measures with respect to a certain energy-like function $E$ to be defined below. Recently, the equilibrium measures were proved to be related to steady states in thermostated systems.\footnote{For example, a Gaussian thermostat \cite{Thermostat} is a collection of particles subject to a non-conservative force $\bf F$ and a constraint
\be
\alpha=  \frac{\sum_j \left( {\bf F}({\bf q}_j)  - \frac{\partial V({\bf q}_i)}{\partial {\bf q}_i}\right) \cdot {\bf p}_j}{ \sum_j {\bf p}_j^2}
\ee
with non-Hamiltonian equations of motions given by 
\be
\frac{d{\bf p}_i}{dt} = - \frac{\partial V({\bf q}_i)}{\partial {\bf q}_i}+{\bf F}({\bf q}_i) - \alpha {\bf p}_i\;\;\;\; \text{and}\;\;\;\;\frac{d{\bf q}_i}{dt} = \frac{{\bf p}_i}{m},
\ee
such that the total kinetic energy remains constant $\sum_j {\bf p}_j^2=\text{const}$.}

Although the equilibrium measure seems to be the most physical it is by no means unique. The situation is completely analogous to the equilibrium statistical mechanics where the Gibbs measure is defined only once the energy function is specified. This is also not a unique choice and under certain circumstances  other measures with respect to other constraints can be more physically relevant (e.g. grand canonical ensemble). The general rule for finding an appropriate measure is given by the MaxEnt (maximal entropy) principle proposed by Jaynes \cite{MaxEnt}. It says that for any given set of constraints on a system or for a given knowledge about the system, the probability measure, which best represents the state of knowledge, is the one with largest entropy.

What is, however, unique about the equilibrium measure $\mu_+$ is that it is the only measure which is a zero-noise limit of small perturbations around deterministic trajectories \cite{Reviews}. In other words, if we slightly perturb our deterministic evolution and take the perturbation to zero, then the equilibrium measure is the only measure which converges to itself. Thus, if we are to construct a measure which respects the quantum-classical correspondence principle then $\mu_+$ might be the only choice within a framework of dynamical systems.\footnote{A possible connection of the equilibrium measures to quantum gravity was expressed in Ref. \cite{QuantumGravity}.} To make the above statement more precise a full quantization of the cosmological systems must be carried out which proved to be a difficult task, although a number of recent attempts have been made to advance our understanding of quantum mechanics on cosmological scales\cite{QuantumMechanics}.

\subsection{Variational principle}

To study the statistical properties of non-Hamiltonian systems, a thermodynamic formalism was developed with many ideas borrowed from the equilibrium statistical mechanics, but one very important difference. In the conventional statistical mechanics we are usually interested in states, when in dynamical system the key role is played by a time-ordered collection of states or by trajectories. Thus, it is convenient to think of time as a thermodynamical volume which is a conjugate variable to the so-called topological pressure. The topological pressure of a given energy-like function $E({\bf x})$ is defined as
\be
p(\beta E) = \lim_{t\rightarrow\infty} \frac{1}{t}  \log {\cal Z},
\ee
where 
\be
{\cal Z} =  \int d{\bf x}(0) e^{ -\beta \int_0^t E({\bf x}(\tau))d\tau}
\ee
is a dynamical partition function.\footnote{There is no $1/\beta$ in the definition of the topological pressure which might cause some confusion whenever $\beta \neq 1$, but it seems to be a standard convention in the dynamical systems literature.} For $\beta E=0$ the topological pressure (known as topological entropy) is equal to the rate of growth of the number of topologically distinguishable trajectories and for $\beta E \neq 0$ these trajectories are also weighted by $\exp(-\beta E)$.  

The energy-like function $E_+$ which corresponds to the forward equilibrium measures $\mu_+$ is given by a sum of local Lyapunov exponents $\chi_i$ (defined as local rates of separation of nearby trajectories) over directions corresponding to only positive (global) Lyapunov exponents $\lambda_i$ (defined as rates of separation of nearby trajectories in the limit of infinite times), i.e.
\be
E_+({\bf x}) = \sum_{\lambda_i>0} \chi_i({\bf x}).
\ee
Similarly, the energy-like function $E_-$, which corresponds to the backward equilibrium measure $\mu_-$, is given by a sum over negative Lyapunov exponents or by a sum over positive Lyapunov exponents of a time-reversed system,
\be
E_-({\bf x}) = - \sum_{\lambda_i<0} \chi_i({\bf x}).
\ee
For Hamiltonian systems the Lyapunov exponents come in conjugate pairs (i.e. $E_+({\bf x}) = E_-({\bf x})$) and the two equilibrium measures (backward and forward) are identical, i.e. $\mu_+=\mu_-$.

Another important quantity is the Kolmogorov-Sinai entropy defined for any given time-invariant measure $\mu({\bf x})$ using the Shannon entropy formula on a space of trajectories. For a discrete time dynamical system $T: X \rightarrow X$ the entropy is defined as\footnote{${\cal C} =\{C_1, C_2,....,C_n\}$ is a finite partition of $X$ if $X = \cup_{i=1}^n C_i$ and $C_j \cap C_k = \emptyset$ for $j\neq k$.}
\be
S_\mu \equiv \text{sup}\{  \lim_{t\rightarrow\infty} \frac{1}{t}  S_\mu(\vee_{k=0}^{t-1} T^{-k}({\cal C}) ) : {\cal C}\; \text{is a finite partition of}\; {X} \}
\ee
where
\be
S_{\mu}({\cal C}) = -  \sum_{i=1}^n \mu(C_i) \log(\mu(C_i)).
\ee
The union of two partitions (${\cal C} =\{C_1, C_2,....,C_n\}$ and  ${\cal D} =\{D_1, D_2,....,D_m\}$) is defined as ${\cal C} \vee {\cal D} \equiv \{ C_i \cap D_j :  C_i \in {\cal C}, D_j \in {\cal D}  \}$. The Kolmogorov-Sinai entropy (pre unit time) can also be defined for a continues time process \cite{Reviews}. Intuitively $S_\mu$ quantifies the significance of long periodic orbits with respect to a given measure $\mu$, and should not be confused with a thermodynamic entropy on states.

We are now ready to state one of the two most important results of the  thermodynamic formalism - a variational principle.\footnote{The variation principle is analogous to the Gibbs variational principle which defines the equilibrium state of a system by minimizing its free energy. Thus, it might be helpful to think of $\int E_+({\bf x}) \mu({\bf x}) d{\bf x}- S_\mu/\beta$ as a dynamical free energy per unit time.} It says that 
\be
p(\beta E_+) = \sup\{ S_\mu  - \beta \int E_+({\bf x}) \mu({\bf x}) d{\bf x} : \mu \in {\cal M}  \}.
\ee
where the extremum is realized for $\mu=\mu_+$. The variational principle allows us to calculate the topological pressure 
\be
p(\beta E_+) = S_{\mu_+} - \beta \int E_+({\bf x}) \mu_+({\bf x}) d{\bf x}
\ee
as well as the equilibrium measure, corresponding to $\beta=1$, 
\be
\mu_+({\bf x}(0)) \propto \exp\left(-\int_0^t E_+({{\bf x}(\tau)}) d\tau\right)
\ee
from a spectrum of Lyapunov exponents. For closed systems without any absorbing sates the topological pressure vanishes, $p(E_+) =0$ and for open systems $p(E_+) =-\gamma$, where $\gamma$ is the escape rate of trajectories.

\subsection{Absorbing states}

In the context of eternal inflation it will be useful to study dynamical systems whose phase space trajectories can either reproduce or escape. The escape of trajectories is easier to understand when a system with absorbing states is followed forward in time. Similarly, if the system with escape is followed backward in time then it would seem as though  the forward trajectories reproduce. Thus, it is convenient to define the reproducing states as absorbing states of a time-reversed system. Of course, if we are only interested in the time-invariant measures, then none of the forward nor backward trajectories would ever escape and the existence of absorbing and reproducing states is not directly observable. 

The simplest dynamical systems with absrobing states are called cookie-cutters. Cookie-cutters are defined by a discrete map $T$ from a union of disjoint subsets $A_i \subset [0,1]$ to the entire unit interval such that $[0,1] - \cup A_i \neq \emptyset$. For example,  
\bea 
T(x) = \begin{cases} 3x\;\;\;&\text{if}\;\;\; x\in [0,1/3] \\
 2 x- 1\;\;\;&\text{if}\;\;\; x\in [1/2, 1].
\end{cases}
\eea
On each iteration the map of an open interval $(1/3,1/2)$ is undetermined which represents a terminal or absorbing state. The set of all points that are never mapped to $(1/3,1/2)$  is $A\equiv\sup\{Y | Y=T(Y)\}$. The set $A$ is a fractal whose Liouville (or more precisely Lebesgue) measure is zero (i.e. $\mu_L(A)=0$), but one could still ask whether it is possible to construct an equilibrium time-invariant measure on $A$.

The space of all time-invariant measures is very large (e.g. $\mu(x) = \delta(x-0)$, $\mu(x) = \delta(x-1/5)/2+\delta(x-3/5)/2$, etc.), but the equilibrium measure  $\mu_+$ is a unique late time limit of an arbitrary (continuous with respect to $\mu_L$) distribution,  i.e.
\be
\mu_+(x) \propto \exp\left (- \beta \sum_{i=0}^{t-1} E_+(T^i(x)) \right ),
\ee
where $E_+(x) = \log \left | \left [\frac{dT}{dy}\right ]_{x} \right |$ is the only positive Lapunov exponents at $x$. Although $\mu_+(x)$ is defined precisely for infinite $t$ one can study a coarse-grained partition function for a fixed $t$ by summing over periodic orbits:
\be
{\cal Z} = \sum_{T^t(x)=x} \exp\left (- \beta \sum_{i=0}^{t-1}  \log \left | \left [\frac{dT}{dx}\right ]_{T^i(x)} \right | \right ).
\ee
Then, the topological pressure is
\be
p = \lim_{t\rightarrow \infty} \frac{1}{t} \log {\cal Z} = \lim_{t\rightarrow \infty} \frac{1}{t} \log (3^{-\beta} + 2^{-\beta})^t= \log (3^{-\beta} + 2^{-\beta}).
\ee
According to Bowen \cite{Bowen}, the vanishing topological pressure $p=0$ implies that $\beta\approx 0.7879$ is the fractal dimension of $A$, but the most physical choice is given by the equilibrium measure with $\beta=1$.

\subsection{Fluctuation theorem}

In dynamical systems literature, the entropy production rate  had been identified with (minus) the phase space contraction rate,
\be
e({\bf x}) = -  \frac{\partial}{\partial {\bf x}} \cdot  {\bf v}({\bf x}).
\ee 
whose average with respect to the equilibrium measures $\mu_+$ is non-negative, i.e.
\be
 - \int \frac{\partial}{\partial {\bf x}} \cdot  {\bf v}({\bf x}) d\mu_+({\bf x}) \geq 0,
\ee 
and is strictly positive for dissipating systems. In our notations the local entropy production rate is given by a sum of all local Lyapunov exponents:
\be 
e({\bf x})   \equiv  \sum_i \chi_i({\bf x}) =  E_+({\bf x}) - E_-({\bf x}),
\ee
where the expressions for both $E_+({\bf x})$ and $E_-({\bf x})$ in the context of eternal inflation will be derived in the following section. Alternative calculations of the entropy production during inflation using stochastic methods is described in Refs. \cite{StochasticEntropy}.

Since any physical system is usually observed only for a finite period of time, one might want to define a finite time average as
\be 
e_T({\bf x}(0)) \equiv \frac{1}{T}\int_{0}^{T}e({\bf x}(t)) dt
\ee
and study the properties of the probability distribution $P(e_T)$. This problem was analyzed in the context of Anosov systems, where it was found that 
\be
\log\left(\frac{P(e_T)}{P(-e_T)}\right) =  e_T T
\label{Eq:FluctuationTherom}
\ee
is exactly linear with no higher order terms for an arbitrarily large $e_T$. This relation is the second of the two most important results of the thermodynamic formalism known as the fluctuation theorem  \cite{FluctuationTheorem, FluctuationProve}. The symmetry of the distribution $P(e_T)$  (sometimes called Gallavotti-Cohen symmetry) involves the statistics of very atypical fluctuations and was first discovered in numerical simulations \cite{FluctuationTheorem}. Later, the fluctuation theorem was also proved analytically for Anosov systems \cite{FluctuationProve}. 

However, if we assume that the chaotic hypothesis holds for a dynamical system of eternal inflation  (i.e. eternal inflation can be regarded as an Anosov system), then we should also expect to see the symmetry described by Eq. (\ref{Eq:FluctuationTherom}) in primordial fluctuations. Note, that this symmetry is more than just Gaussianity, but involves the statistics of very improbable events. On a positive side, the Cosmic Microwave Background experiments allow us to retrieve information about separate (i.e. causally disconnected) trajectories by simply looking at different directions on the sky. This is certainly an advantage compared to other non-equilibrium systems  such as thermostats \cite{Thermostat} where the fluctuation relation is being tested. For example, one can divide the CMB sky into $N$ equal regions and use the CMB data to estimate what could have been the entropy production $e({\bf x})$ during inflation on each of the corresponding trajectories separately. Then, according to the fluctuation theorem, the distribution $P(e_T)$ must have a symmetry described by Eq. (\ref{Eq:FluctuationTherom}) which can always be verified for a sufficiently large $N$ and sufficiently small $T$.

\section{Eternal inflation}\label{SecEternal}

The main objective of this section is to construct the time-invariant equilibrium measures of eternal inflation using dynamical systems, but before we proceed it is instructive to highlight the main properties of such measures within a more familiar stochastic approach. Since all time-invariant measures, $\mu \in {\cal M}$, are defined on a space of infinite (or very large) trajectories, the relevant observables are the trajectories.\footnote{A possibility of defining measures on trajectories was already expressed in Refs. \cite{TrajectoriesStochastic}.} As was emphasized above, this might potentially solve the problem of defining the relevant cosmological observables. For example, one can show that various paradoxes \cite{GeocentricCosmology, GuthVanchurin, Paradoxes} can be resolved whenever the probabilities of cosmological observations are defined on a space of trajectories. In addition, the time-invariant measures provide a simple solution to the problem of initial conditions which does not exist for infinite trajectories.\footnote{The measures on only non-singular infinite trajectories were also discussed in Ref. \cite{Page} in the context of homogeneous cosmologies.} In eternal inflation literature such trajectories are usually neglected on the grounds of zero measure (with respect to comoving volume), but because of their infinite lengths one might also argue that any infinite trajectory is infinitely more probable than any finite trajectory. Clearly, there is an order of limits issue that we are going to discuss next.

\subsection{Order of limits}

Consider a stochastic mode of eternal inflation with absorbing (or terminal) states. The only relevant, for our considerations, parameter is the decay rate per unit time $\gamma$ to one of the absorbing states. Our task is to define a measure $\mu(T)$ on a measurable space of trajectories parametrized by their duration $T$ before the final transition to an absorbing state. There are two factors that might go into $\mu(T) \propto w_{\text{evolution}}(T) w_{\text{observation}}(T)$. First of all, the trajectories could be weighted by their probabilities with respect to some Markovian evolution operator, $w_{\text{evolution}}(T)$. In addition, the trajectories could also be weighted by some monotonic function $w_{\text{observation}}(T)$ such that $w_{\text{observation}}(\infty)=\infty$. The first factor can be argued for using, for example, semiclassical methods \cite{StarobinskyStochastic, LocalStochastic}, and the second factor can be argued for using, for example, anthropic principle \cite{Anthropics} since the longer trajectories intersect more observers that could observe them.

Now, if we compare forward trajectories of length $T$, with trajectories of infinite length, then from the point of view of evolution 
\be
\frac{w_{\text{evolution}}(T)}{w_{\text{evolution}}(\infty)} = \frac{\exp(-\gamma T)}{\exp(-\gamma \infty)} = \infty,
\ee
but from the point of view of observations
\be
\frac{w_{\text{observation}}(T)}{w_{\text{observation}}(\infty)} = 0,
\ee
where the exact form of  $w_{\text{observation}}(T)$  is not important. To make sense of $\frac{\mu(T)}{\mu(\infty)} = \infty \times 0$ we can introduce two separate cut-offs,
\be
\frac{\mu(T)}{\mu(\infty)} = \lim_{b \rightarrow \infty}  \lim_{a \rightarrow \infty} \frac{w_{\text{evolution}}(T)}{w_{\text{evolution}}(b)}   \frac{w_{\text{observation}}(T)}{w_{\text{observation}}(a)},
\label{Eq:OrderOfLimits}
\ee
where the order of limits is not specified a priori.  The standard choice is to first take $b \rightarrow \infty$ (or to take both limits simultaneously $a=b \rightarrow \infty$), which usually gives a divergent answer, i.e. $\frac{\mu(T)}{\mu(\infty)} = \infty$, or a zero measure for infinite trajectories. Such ordering of limits is known to give answers that depend on ether initial conditions (for local stochastic measures), or cutoffs (for global stochastic measures).  Another alternative is to first take $a \rightarrow \infty$ which would yield a completely different answer, i.e. $\frac{\mu(T)}{\mu(\infty)}  = 0$. According to the dynamical systems approach the latter ordering procedure is a lot more natural and corresponds to the time-invariant measures. 

The situation is very similar in a negative time direction. It is well known that all but a measure-zero (with respect to comoving volume) of trajectories are past-incomplete \cite{BGV} and that is why all past-complete trajectories are usually neglected. However, if the relevant measurable space is a space of trajectories, then the measure on trajectories depends on the order of limits as in Eq. (\ref{Eq:OrderOfLimits}) and the infinite past-complete trajectories must not be overlooked. In fact, such trajectories are the only trajectories which acquire a finite weight with respect to the ordering procedure suggested by the dynamical systems approach. However, in the models, where the past-complete infinite trajectories are strictly forbidden the time-invariant measures would still be supported on only future-infinite trajectories.

In a stochastic picture the time-invariant measures might be thought of as fractal measures defined on only eternal geodesics \cite{LocalStochastic} in the limit when the cut-off $a$ is taken to infinity and $b$ remains large, but finite. Of course, for any finite value of $a \gg b$ the measure assigns a non-vanishing weight to all trajectories within a small neighborhood around each eternal geodesic. It might be tempting to conclude that the probability to observe a terminal vacua (i.e.  AdS vacua) is extremely small, if not identically zero. This is a very nice prediction which is in agreement with the observed positive value of the cosmological constant. On the other hand, if there would be a mechanism (perhaps quantum) to resolve the black-hole singularities by recycling local trajectories back to eternal inflation, then the system would not contain any absorbing states and the above conclusion would certainly change. 

\subsection{Backward measure} 

We are now ready to switch to a more formal measure-theoretic discussion of eternal inflation. To warm up, we start with a construction of a backward (in time) equilibrium measure for a single scalar field inflation. The objective is to describe the dynamics of the field from the point of view of a local observer moving along a time-like geodesic. The background equations of motion are given by
\be
\frac{d{\pi}}{dt} = - \partial_\varphi V + 3H {\pi}
\ee
and
\be
\frac{d{\varphi}}{dt}= \pi
\ee
where $\varphi$ and  $\pi$  are the position and momentum coordinates. In Minkowski space $H=0$ and the dynamics is described by a time-independent Hamiltonian leading to an incompressible flow of the nearby phase space trajectories, but during inflation $H\neq 0$ and the flow becomes compressible. More precisely, the Hubble friction introduces a single negative local Lyapunov exponent $\chi_i$ of a forward evolution or a single positive local Lyapunov exponent of a backward evolution. If this exponent has the same sign as the corresponding (global) Lyapunov exponent $\lambda_i$, then 
\be
E_-({\bf x})  =  - \sum_{\lambda_i<0} \chi_i({\bf x}) \approx 3H({\bf x}),
\label{Eq:BackwardRate}
\ee 
where ${\bf x}(t)\equiv\{\pi(t), \varphi(t)\}$. In fact, every field contributing to a quasi de Sitter expansion by forming a condensate is likely to contribute to a sum of the negative local Lyapunov exponents, but, for the time being, we assume that there is only a single scalar field that drives inflation. Note that all non-inflating fields would also have negative local Lyapunov exponents $\chi_i$ during inflation, but their relation to signs of the corresponding (global) Lyapunov exponents $\lambda_i$  is not direct, and we will assume that on average they do not contribute to the compressibility of the flow.

Since the Hubble friction introduces only negative Lyapunov exponents it plays a central role for constructing the equilibrium measure on backward trajectories, i.e.
\be
\mu_-({\bf x}(0)) \propto \exp\left( {- \int_0^t E_-({\bf x}(\tau)) d\tau}\right) =  \exp\left( {- \int_0^t 3H({\bf x}(\tau)) d\tau}\right),
\label{Eq:BackwardMeasure}
\ee 
but has no effect on the equlibrium measure of forward trajectories. This implies that  $E_+({\bf x})\neq E_-({\bf x})$ (at the level of background dynamics) and $\mu_+ \neq \mu_-$ which should not be too surprising. The equilibrium measures on forward and backward trajectories of a non-Hamiltonian system usually differ even though they are defined on the same space of infinite trajectories. In fact this will turn out to be the case for the dynamical system of eternal inflation even when perturbations are taken into account. 

\subsection{Forward measure}

At the level of background dynamics the nearby phase space trajectories do not expand and one must go beyond the homogeneous limit in order to derive the equilibrium measures on forward trajectories. In this limit the quantum effects cannot be ignored and one often employes the semiclassical tools to study quantum fluctuation generated during inflation. It is well known that the semiclassical analysis gives rise to a stochastic picture which is sufficient for modeling inflation using diffusion Eqs.  (\ref{Eq:GlobalStochastic}) and (\ref{Eq:LocalStochastic}), but for the methods developed in the previous section to be useful we should also learn how to extract the microscopic properties. In particular, it is desired to map the microscopic parameters, such as Lyapunov exponents, to the macroscopic parameters, such as transport coefficients. This would enable us to estimate the equilibrium measures on forward trajectories and to study their properties.

A number of different ideas had been put forward to address this issue, but perhaps the simplest of all is the escape-rate formalism \cite{StochasticDeterministic}.   The idea is to express the rate of escape $\gamma$ from a given phase space neighborhood of size $L$ using thermodynamic formalism and then to equate it to the escape rate calculated using a diffusion equation, i.e.
\be
\gamma = - p(E_+)=\int E_+({\bf x}) \mu_+({\bf x}) d{\bf x} - S_{\mu_+} =  \left(\frac{\pi}{L}\right)^2 D
\label{Eq:Escape-rate}
\ee
where $S_{\mu_+}$ is the Kolmogorov-Sinai entropy of the neighborhood.  The semiclassical analysis \cite{StarobinskyStochastic} suggests that the evolution is described by Eq. (\ref{Eq:LocalStochastic}) with diffusion coefficient given by
\be
D   = \frac{H^3}{8 \pi^2},
\label{Eq:DiffusionCoefficient}
\ee
where we ignore the problem of factor ordering. Then, according to Eqs. (\ref{Eq:Escape-rate}) and (\ref{Eq:DiffusionCoefficient}), the sum of positive local Lyapunov exponents is likely to scale linearly with diffusion coefficient, i.e. 
\be
E_+({\bf x}) \propto H({\bf x})^3.
\ee 
This is our best guess of what the underlying microscopic properties of the system should be based on the semiclassical methods \cite{StarobinskyStochastic} and on the escape-rate formalism \cite{StochasticDeterministic}. Of course, one would want to go beyond the semiclassical theory to confirm (or disprove) the linear dependence of microscopic Lyapunov exponents on macroscopic diffusion coefficients.  

In addition to diffusion, the fluctuations are constantly stretched by cosmological expansion which gradually reduces the effect of any particular mode (along a given local trajectory) on the Lyapunov spectrum. This can be captured by an additional factor $\propto a({\bf x})^3$ in the expression for $E_+({\bf x})$, where $a({\bf x})$ is a local scale factor which describes the local FRW geometry. Then the final expression for the sum of positive Lyapunov exponents is
\be
E_+({\bf x}) \propto H({\bf x})^3  a({\bf x})^3
\label{Eq:ForwardRate}
\ee 
and the corresponding equilibrium measure is
\be
\mu_+({\bf x}(0)) \propto \exp\left( {- \int_0^t E_+({\bf x}(\tau))  d\tau}\right) = \exp\left( {- \beta\int_0^t   H({\bf x}(\tau))^{3} {a}({\bf x}(\tau))^3 d\tau}\right),
\label{Eq:ForwardMeasure}
\ee
where $\beta$ is yet undetermined constant. We would like to stress that the above equation does not contain all of the quantum effects (e.g. quantum tunnelings), but only provides an estimate of the effect of linear inflationary perturbations on the equilibrium measure of forward trajectories due to positive Lyapunov exponents.

\subsection{Effect of horizons}

In the dynamical systems of inflation discussed so far the local phase space trajectories did not escape nor reproduce, but a generic dynamical system may contain both absorbing and reproducing (i.e. absorbing states of a time-reversed system) states. For such systems the rate of escape of forward trajectories is usually given by the sum of positive Lyapunov exponents (\ref{Eq:ForwardRate}). Then the contribution to the escape rate from $N$ scalar fields is given by 
\be
E^{\text{phase-space}}_+({\bf x})  \approx  \sum_{i=1}^{N}  \beta_i H({\bf x})^{3} a({\bf x})^3.
\label{Eq:ForwardLocal}
\ee
Roughly speaking, the larger the rate of a phase space expansion the easier it is for a given trajectory to escape or to hit an absorbing state if such a state exists. Similarly, the rate of reproduction of local trajectories is given by the rate of escape of backward trajectories or by minus the sum of negative Lyapunov exponents (\ref{Eq:BackwardRate}). For inflation driven by $N$ scalar field the rate is 
\be
E^{\text{phase-space}}_-({\bf x})  \approx  \sum_{i=1}^{N} 3 H({\bf x}) =  3 N H({\bf x}).
\label{Eq:BackwardLocal}
\ee
This is a phase-space picture.  

At the same time, eternal inflation can be described from a physical-space point of view where the local trajectories reproduce or escape due to the presence of cosmological and black hole horizons. According to the physical-space picture the local trajectories reproduce with the following rate 
\be
\frac{d}{dt} \log\left(\frac{V_0 a^3}{H^{-3}}\right) = 3 \;H({\bf x}) + 3\frac{\dot{H}({\bf x})}{H({\bf x})}.
\label{Eq:Rate}
\ee
where $V_0 a^3/H^{-3}$ is the number of independent local trajectories at time $t$ inside of comoving volume $V_0$. The second term of Eq. (\ref{Eq:Rate}), averaged over periodic orbits, is exactly zero, but the first term gives a non-zero contribution to the reproduction rate, 
\be
E^{\text{physical-space}}_-({\bf x})  =  3 \;H({\bf x}).
\label{Eq:BackwardGlobal}
\ee
Roughly speaking, this is the rate with which local observers fall out of causal contact with each other and start to follow their own local trajectories. 

In addition to reproduction, the physical-space picture suggests that the local trajectories might eventually escape into black hole singularities whenever the energy density contrast during inflation is of order one or larger. This is usually the case for inflationary perturbations generated above the self-reprodcution scale, if they are not stretched out by dark energy, but might occasionally happen on any scale if the classical drift of a scalar field is of the same order or smaller than a given quantum jump, i.e $\delta\varphi_{\text{class}} \lesssim \delta\varphi_{\text{quant}}$. For Gaussian fluctuations (below self-reproduction scale) one finds that the corresponding rate of escape is, 
\be
E^{\text{physical-space}}_+({\bf x}) \sim \frac{H({\bf x})^3}{\partial_\varphi H({\bf x})}.
\label{Eq:ForwardGlobal}
\ee
Of course, the above equation is only valid towards the end of inflation so that the fluctuations have time to reenter horizon and the local observers have time to escape into the singularity before the cosmological constant starts to dominate.

To apply the methods developed in this paper to an arbitrary model of eternal inflation one might want to describe the effects of horizons described by Eqs. (\ref{Eq:BackwardGlobal}) and (\ref{Eq:ForwardGlobal}) in a language of dynamical systems. This can be accomplished by generalizing the energy-like functions $E_-$ and $E_+$ to include both the phase space and physical space contributions, i.e.
\be
E_-({\bf x})  = E^{\text{phase-space}}_-({\bf x})+E^{\text{physical-space}}_-({\bf x}) 
\label{Eq:BackwardSum}
\ee 
and
\be
E_+({\bf x}) =E^{\text{phase-space}}_+({\bf x})+E^{\text{physical-space}}_+({\bf x}).
\label{Eq:ForwardSum}
\ee
Depending on a model of eternal inflation the horizons may or may not change significantly the corresponding equilibrium measures of Eqs. (\ref{Eq:BackwardMeasure}) and (\ref{Eq:ForwardMeasure}).

\section{Summary of results}\label{SecResults}

The dynamical systems approach to non-equlbrium statistical mechanics is a viable alternative to the stochastic approaches with a wide range of applications. For cosmological systems the approach is introduced here for the first time, but it is based on a well known mathematics developed over the last few decades. The main motivation to study something new was to solve the cosmological problems of entropy, measure, observables and initial conditions. We do not claim that the solution presented here is unique, but we do claim that none of the previously proposed scenarios solve all of these problems simultaneously. Below we list the most popular frameworks\footnote{Other promising holographic approaches which are currently under development include dS/CFT \cite{DSCFT}, FRW/CFT \cite{FRWCFT}, dS/dS \cite{DSDS}, HST \cite{HST} and Holographic Multiverse \cite{HVG}.}  and their problems:\footnote{``No'' means that the respective problem can be solved or avoided at least for some systems, and ``Yes'' means that, to our knowledge, the problem cannot be solved nor avoided within a given framework.}
\newline
\newline
\begin{tabular}{ c |c|c|c|c }\hline
Framework & Entropy & Measure   & Observables & Initial Conditions  \\
\hline
Quantum Cosmology \cite{QuantumCosmology, HartleQuantum}& No & Yes & No & Yes  \\
Local Stochastic \cite{StarobinskyStochastic, LocalStochastic, BoussoStochastic} & No & No &  Yes & Yes \\
Global Stochastic \cite{LindeStochastic,VilenkinStochastic} & No & Yes & Yes & No \\
Stationary Stochastic \cite{StationaryStochastic} & No & Yes & No & No \\
Geocentric Cosmology \cite{GeocentricCosmology} & No & Yes & No & No \\
Hamiltonian Systems \cite{EntropyProblem} & Yes & No  & No & No \\
Dynamical Systems & No & No & No & No \\
\hline
\end{tabular}
\newline
\newline
\newline
Perhaps the most interesting result about the new approach is that, if correct, it gives us a hope to derive the cosmological predictions from the dynamics itself without a need to postulate any additional rules (e.g. measure, initial conditions, space of obervables etc.). This would be a truly dynamical solution to the existing cosmological problems.

We conclude with a critical summary of the main results:

{\it 1) Problem of observables.} Many cosmological measures, defined over a space of local states, are known to suffer from serious logical inconsistencies. Although, the probability spaces over local states are natural within a stochastic approach, it is not the case for the dynamical systems approach developed in this paper. The new framework suggests that the relevant observables should be local trajectories and the measures should be defined over the space of trajectories instead of states. Without going into details we mention that one might still run into philosophical issues (e.g. terminal states have zero probability) whose implications remain to be better understood. 

{\it 2) Entropy problem.} The Liouville measure is the most physical measure for a finite Hamiltonian system which is known to suffer from the entropy problem. A solution proposed here involves a generalization of a purely Hamiltonian dynamics to include absorbing  and reproducing states. For such systems the most physical measures are not the Liouville measures, but more general equilibrium measures. This does not mean that a general dynamical system would never be in a conflict with observations, but it does mean that such problems can be avoided. 

{\it 3) Measure problem.} There are many ``solutions'' to the entropy problem using stochastic processes (i.e. local, global, stationary), but their major drawback is that they always introduce new problems such as the problem of measure or the problem of initial conditions. This is not the case for the dynamical systems approach whose equilibrium measures are uniquely defined by a variational principle. Of course, for some dynamical systems close to a dynamical phase transition the equilibrium measure might still be degenerate and such critical systems certainly deserve a closer examination.\footnote{The dynamical phase transitions usually occurs in non-hyperbolic dynamical systems and should not be confused with phase transitions in statistical mechanics.} 

{\it 4) Problem of initial conditions.} The equilibrium measures are only supported on infinite trajectories to the past as well as to the future. For such measures the initial conditions are irrelevant by construction. However, even if one demands to start with a distribution (continuous with respect to the Liouville measure) the dynamical system would eventually forget its initial state, similarly to what happens in finite Hamiltonian systems. 

{\it 5) Fluctuation theorem.} The dynamical systems approach would not be very useful without the chaotic hypothesis. The hypothesis is a natural generalization of the ergodic hypothesis and is often assumed for analysis of sufficiently chaotic systems. One of the results that follows immediately is a symmetry described by the fluctuation theorem which could in principle be observable in the Cosmic Microwave Background radiation where the nearby local trajectories can be compared side by side. This involves the analysis of very improbable fluctuations which is a challenging task.

{\it Acknowledgments.} The author is grateful to Alan Guth, Andrei Linde, Mahdiyar Noorbala and Alexander Vilenkin for very useful discussions and comments on the manuscript. The work was supported in part by NSF grant PHY-0756174.

\end{document}